\documentclass[aps,prb,showpacs]{revtex4}
\usepackage{epsfig}
\usepackage{amsmath}
\newcommand{\bs}{\begin{subequations}}
\newcommand{\es}{\end{subequations}}
\newcommand{\be}{\begin{equation}}
\newcommand{\ee}{\end{equation}}
\newcommand{\bea}{\begin{eqnarray}}
\newcommand{\eea}{\end{eqnarray}}
\newcommand{\bd}{\begin{displaymath}}
\newcommand{\ed}{\end{displaymath}}
\newcommand{\ba}{\begin{array}}
\newcommand{\ea}{\end{array}}
\newcommand{\nn}{\nonumber}

\newcommand{\Del}{\Delta}
\newcommand{\del}{\delta}
\newcommand{\Gam}{\Gamma}

\newcommand{\al}{\alpha}

\newcommand{\eps}{\epsilon}

\newcommand{\ua}{\uparrow}
\newcommand{\da}{\downarrow}

\newcommand{\til}{\tilde}
\begin{document}

\title{\bf Analytical solution of the time evolution of an entangled electron spin pair in a double quantum dot nanostructure}

\author{
M. Blaauboer
}

\affiliation{Kavli Institute of Nanoscience, Delft University of Technology,
Lorentzweg 1, 2628 CJ Delft, The Netherlands}
\date{\today}

\begin{abstract}
Using master equations we present an analytical solution of the time evolution of an entangled electron spin pair which can occupy 36 different quantum states in a double quantum dot nanostructure. This solution is exact given a few realistic assumptions and takes into account relaxation and decoherence rates of the electron spins as phenomenological parameters. Our systematic method of solving a large set of coupled differential equations is straightforward and can be used to obtain analytical predictions of the quantum evolution of a large class of complex quantum systems, for which until now commonly numerical solutions have been sought.
\end{abstract}

\pacs{73.63.-b, 02.50.Ga, 03.65.Yz}
\maketitle

\section{Introduction}

Master equations are used to describe the quantum evolution of a physical system interacting with some ``reservoir''~\cite{blum96}, and have been applied to a wide variety of physical systems, ranging from two-level atoms in the presence of light fields~\cite{meys99} to solid-state nanostructures such as quantum dots and Josephson junction devices~\cite{makh01}. For simple systems, such as a two-level atom damped by a reservoir consisting of simple harmonic oscillators~\cite{meys992} or an electron in a single or double quantum dot coupled to external leads~\cite{enge01}, the set of master equations that describes the quantum dynamics of the system is small and its solution can be obtained analytically in a straightforward way. If the system is more involved, however, due to the presence of quite a few atomic levels or because the nanostructure is composed of various coherent parts, its quantum state space consists of a large number of quantum states with various coherent and incoherent couplings between them and the analytical solution of the corresponding large set of coupled master equations does not spring to the eye. Hence often a numerical solution is sought~\cite{sara03}. Understanding the quantum evolution of such ``complex'' quantum systems - where ``complex'' refers to a system which is described by a large number of coupled quantum states - has recently become increasingly important, in particular in fundamental research aimed at investigating the dynamic behavior of qubits, the basic building blocks for quantum computation~\cite{raim01}. A large theoretical and experimental effort in various fields, e.g. quantum optics, atomic physics and condensed matter physics, is presently directed
towards investigating possibilities to use two-level systems such as polarized photons, cold atoms, electron spins and superconducting circuits as qubits and finding ways to couple these qubits together. In the latter three systems, one of the major questions involved is how the desired coherent evolution of the system will be affected by coupling to the environment, which is necessary to manipulate and measure the states of the qubits but invariably introduces undesired decoherence of their quantum states. A master equation model of the quantum evolution of one or more qubits interacting with their environment allows one to construct transparent general formulas and is therefore very suitable to give both qualitative and quantitative insight into the dynamics of these complex quantum systems. 

In this paper we present an analytical solution of a large set of coupled master equations that describe the quantum evolution of a particular condensed-matter system, namely the time evolution of an entangled electron spin pair in a double quantum dot nanostructure. Even though our model applies to this specific quantum system, the presented method of solving the master equations is general and can be applied to study the dynamics of many other complex quantum systems. The time evolution of the electron spins is governed by several coherent and incoherent processes, each of which depends on time in a simple way as either oscillatory (cosine) or exponential functions. The solution we obtain shows how these simple ingredients combine to describe the evolution of the entangled spins in a complex nanostructure which consists of several coherent parts. It can be used to predict the occupation probability of all quantum states at any given time and to provide analytical estimates of the important time scales in the problem, such as the time at which decoherence of the entangled pair becomes substantial~\cite{decoherence}. 

The paper is organized as follows. In Sec.~\ref{sec-system} the quantum
nanostructure and the assumptions made are described. Section~\ref{sec-solution}
contains the master equations and their solution, with technical details given in Appendices~\ref{app-A} and \ref{app-B}. A summary of the results and their range of
applicability is presented in Sec.~\ref{sec-discussion}.

\section{The double quantum dot nanostructure}
\label{sec-system}

\begin{figure}[h]
\centerline{\epsfig{figure=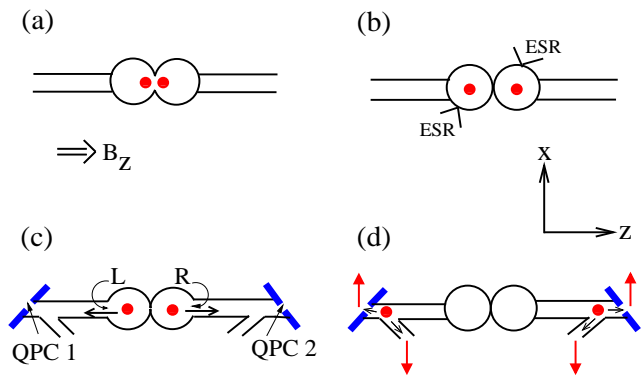,height=5.cm,width=8.5cm}}
\caption{Schematic top view of the double quantum dot nanostructure as
discussed in Sec.~\ref{sec-system}.
}
\label{fig:system}
\end{figure}

The system we consider consists of a double quantum dot
nanostructure, which is occupied by two entangled electron spins and operated
as a turnstile. We studied this system in an earlier paper as a suitable
set-up for the detection of entanglement between electron
spins~\cite{blaa05}. Here we focus on the dynamic evolution of the
electron pair in the system, which is depicted in
Fig.~\ref{fig:system}.

In detail, the structure consists of two adjacent quantum dots in a parallel magnetic field
$B_z \hat{z}$ which are connected to two quantum point contacts (QPCs)
via empty quantum channels. A quantum dot is a small
metallic or semiconducting island, confined by gates and connected to
electron reservoirs (leads) through quantum point contacts. 
If the gates are nearly-closed and form tunnel barriers, the dot is occupied by a 
finite and controllable number of 
electrons which occupy discrete quantum levels, similar to atomic orbitals 
in atoms~\cite{kouw97}. 
In our system, the gate between the two dots is assumed to be initially open 
and the dots are occupied by two electrons~\cite{singlet1}
[Fig.~\ref{fig:system}(a)] in their lowest energy state, the singlet
state~\cite{singlet2}. The gate between the two dots is then
adiabatically closed, so that the electrons become separated and one dot
is occupied by an electron with spin-up and the other by one with spin-down.
The two spins do not interact anymore and are independently rotated by
electron spin resonance (ESR) fields [Fig.~\ref{fig:system}(b)]. 
The latter are oscillating magnetic fields which, if the frequency of 
oscillation matches the energy difference between the two spin-split 
single-electron energy levels, cause coherent rotations of a spin 
between these levels, analogous to Rabi oscillations in a two-level atom. 
After spin rotation, the electrons
are emitted into empty quantum channels by opening gates L and R
[Fig.~\ref{fig:system}(c)] and scattered at quantum point contacts QPC 1
and QPC 2. In a parallel magnetic field and for conductances
$G_{\rm QPC1 (QPC2)} \leq  e^2/h$ these QPC's are
spin-selective~\cite{poto02}, transmitting electrons with spin-up
and reflecting those with spin-down [Fig.~\ref{fig:system}(d)].
The transmitted and reflected electrons are separately detected in
the four exits.

In the next section we analyze the dynamics of the two spins from the
moment they are separated and each occupies one of the two dots,
until both have been detected in one of the four exits. We use a
master equation approach in which the effects of relaxation and 
decoherence are included as phenomenological decay rates~\cite{blum96}. 
The solution presented is exact
under three assumptions:
\begin{itemize}
\item{The time evolution during ESR in the dots is decoupled from the time evolution 
in the channels and exits. Physically, this means that the gates between the dots and
channels are closed during the ESR rotations, so no tunneling occurs out of the dots
during that time. 
}
\item{Once the electrons are in a channel they cannot tunnel back into
the dots, i.e. backreflection of the electrons to the dots during
their journey to the detectors is neglected. This corresponds to ballistic transport
through the channels.
}
\item{Once the electrons are in one of the exits they cannot return to
the channels, i.e. the electrons are immediately detected and absorbed into 
the detectors.
}
\end{itemize}

\section{The master equations and their solution}
\label{sec-solution}

In the set-up as depicted in Fig.~\ref{fig:system} each electron is
assumed to be either in a dot, in a channel or detected. This leads to a
set of 36
possible quantum states represented by a 36$\times$36 density matrix
$\rho(t)$. This set consists of all possible combinations $A\sigma
B\sigma^{\prime}$, with $A$$\in$$\{D,C,X \}$ and
$\sigma$$\in$$\{\ua,\da\}$
indicating resp. the position ($D$$=$dot, $C$$=$channel and $X$$=$exit) 
and the spin direction along $\hat{z}$ of the electron which started out
in the left dot, and $B$$\in$$\{D,C,X \}$ and $\sigma^{\prime}$$\in$$\{\ua,\da\}$ 
representing
the position and spin direction of the electron which started out
in the right dot. The set is given by:
\bea  
&\left\{  D\!\ua\!D\!\ua,\, D\!\ua\! D\!\da,\, D\!\da\! D\!\ua,\, D\!\da\! D\!\da,\, C\!\ua\! D\!\ua,\, C\!\ua\!
D\!\da,\, C\!\da\! D\!\ua,\, C\!\da\! D\!\da,\, D\!\ua\! C\!\ua, \right. & \nn \\
& D\!\ua\! C\!\da,\, D\!\da\! C\!\ua,\, D\!\da\! C\!\da,\, C\!\ua\! C\!\ua,\, C\!\ua\! C\!\da,\, C\!\da\! C\!\ua,\, 
C\!\da\! C\!\da,\, X\!\ua\! D\!\ua,\, X\!\ua\! D\!\da, & \nn \\
& X\!\da\! D\!\ua,\, X\!\da\!D\!\da,\, D\!\ua\! X\!\ua,\, D\!\ua\! X\!\da,\, D\!\da\! X\!\ua,\, D\!\da\! X\!\da,\, 
X\!\ua\! C\!\ua,\, X\!\ua\! C\!\da,\, X\!\da\! C\!\ua,  & \nn \\
& \left. X\!\da\! C\!\da,\, C\!\ua\! X\!\ua,\, C\!\ua\! X\!\da,\, C\!\da\! X\!\ua,\, C\!\da\! X\!\da,\, 
X\!\ua\! X\!\ua,\, X\!\ua\! X\!\da,\, X\!\da\! X\!\ua,\, X\!\da\! X\!\da
\right\}.
\label{eq:states}
\eea
We number the states in set (\ref{eq:states}) by the numbers 1 to 36, so
1=D$\ua$D$\ua$, 2=D$\ua$D$\da$ etc. The states labeled by C and X do not
refer to {\it individual} quantum states in the channels and detectors, since in a 
channel many longitudinal modes exist and the detectors consist of many 
quantum states  which form together a macroscopic state. What is meant by the 
states C and X is the {\it set} of all channel modes resp. of all quantum 
states in the detectors. These states thus describe the probability of an 
electron to occupy any one of these channel modes or detector states. We come back to why 
this definition is useful and appropriate in the paragraph below Eq.~(\ref{eq:decoherenceC}). 
For long times, 
the only states that are occupied are 33-36, in which both electrons have entered 
into an exit and the channels and dots are empty. \\ 
The time evolution of the density
matrix elements
$\rho_{nm}(t)$ is given by the master equations~\cite{blum96}:
\begin{subequations}
\bea
\dot{\rho}_{n}(t) & = & -\frac{i}{\hbar} \left[ {\cal H}(t), \rho(t)
\right]_{nn} + \sum_{m \neq n} \left( W_{nm}\, \rho_{m}(t) - W_{mn}\,
\rho_{n}(t) \right), 
\label{eq:Mastera}\\
\dot{\rho}_{n,m}(t) & = & -\frac{i}{\hbar} \left[ {\cal H}(t), \rho(t)
\right]_{nm} -
V_{nm}\, \rho_{n,m}(t) \hspace{1cm} n \neq m
\label{eq:Masterb}
\eea
\label{eq:Master}
\end{subequations}
\noindent for $n,m$ $\in$ $\{1,\dots,36\}$. The Hamiltonian ${\cal
H}(t)$ describes the
coherent evolution of the spins in the quantum dots due to the ESR
fields and is given by, for
two oscillating magnetic
fields $B_{xL} \cos(\omega t)\, \hat{x}$ and $B_{xR} \cos(\omega t)\,
\hat{x}$ applied to the left and right dots respectively,
\bea
{\cal H}(t) &= &{\cal H}_{0} - \frac{1}{2} g^{*} \mu_{B} \cos(\omega t)
\! \! \! \!
\! \! \! \sum_{\stackrel{M,N \in \{L,R \}}{M\neq N}} \! \! \! \! \! \!
(B_{xM} + \eps
B_{xN}) \bar{\sigma}_{xM}.
\label{eq:Hamiltonian}
\eea
Here ${\cal H}_{0}$ is a diagonal matrix containing the
energies $E_{n}$ ($n=1,\dots,36$) of each state, $g^{*}$ the electron
g-factor, 
$\mu_B$ the Bohr magneton and
$\bar{\sigma}_{xL(R)}$ a 36$\times$36 matrix with elements
$(\bar{\sigma}_{xL(R)})_{ij}=1$ for each pair of states $(i,j)$
that are coupled by the oscillating field $ B_{xL(R)}$ and zero
otherwise. For $g^{*}<0$ the 4$\times$4 upper left 
corner ${\cal H}_{\rm dots}(t)$ of 
${\cal H}(t)$ is then given explicitly as
\bea
{\cal H}_{\rm dots}(t) = 
\left( \ba{cccc}
E_1 & \hbar \Del_{RL} \cos(\omega t) & \hbar \Del_{LR}\cos(\omega t) & 0 \nn \\
\hbar \Del_{RL}\cos(\omega t) & E_2 & 0 & \hbar \Del_{LR}\cos(\omega t) \nn \\
\hbar \Del_{LR}\cos(\omega t) & 0 & E_3 & \hbar \Del_{RL}\cos(\omega t) \nn \\
0 & \hbar \Del_{LR} \cos(\omega t)  & \hbar \Del_{RL}\cos(\omega t) & E_4 
\ea \right),
\eea
with  $E_1 = 2 E_{\ua} + E_C$, $E_2 = E_3 = E_{\ua} + E_{\da} + E_C$ and
$E_4 = 2 E_{\da} + E_C$ in terms of the single-particle energies
$E_{\ua}$ and $E_{\da}$ and the charging energy $E_C=e^2/C$, where $C$ is the total 
capacitance of the quantum dot (assumed to be equal for both dots), $\Del_{RL} \equiv
\Del_{R} + \del_{L}$ and $\Del_{LR} \equiv \Del_{L} + \del_{R}$ with
$\Del_{R(L)} \equiv \frac{|g^{*}| \mu_B B_{xR(L)}}{2 \hbar}$
and $\del_{R(L)} \equiv \frac{\eps |g^{*}| \mu_B
B_{xR(L)}}{2 \hbar}$. The parameter $\eps$, with $0 \leq \eps<1$,
represents
the relative reduction of the field which is applied to one dot
at the position of the spin in the other dot~\cite{blaa05}. The remaining 32$\times$32
part of the matrix ${\cal H}(t)$ is diagonal and equal to ${\cal H}_{0}$,
since the ESR fields are applied when both electrons are located in a dot and the 
quantum channels do not contain any electrons whose spin might otherwise also be 
rotated by these fields.
\newline
Turning to the transition rates $W_{nm}$ (from state m to n) in
Eqs.~(\ref{eq:Mastera}), we distinguish between two
kinds of transitions: 1) spin-flip transitions between two quantum
states
that differ by the direction of one spin only
and 2) tunneling (without spin-flip) between quantum states that
involve adjacent parts of
the system, i.e. from dot to channel and from channel to exit.
The latter are externally controlled by opening and closing
the gates between the dots and channels. The
former are modeled by the phenomenological rate $1/T_{1,\al} \equiv
W_{\al \ua \da} + W_{\al \da \ua}$ with $\al \in \{D,C \}$ for spin
flips in a dot or channel.
Here the $W's$ depend on the Zeeman energy $\Del E_Z \equiv |g^{*}| \mu_B B_z$ 
and temperature $T$ via detailed balance $W_{\al \ua
\da}/W_{\al \da \ua}= e^{\Del E_Z/k_{B} T}$ 
, so
that
\be
W_{\al \ua \da\, (\da \ua)} =  \frac{1}{T_{1,\al}}\, \frac{1}{1 +
e^{- (+)\Del E_Z /k_{B} T}}, \hspace*{1.cm} \al \in \{D,C \}.
\label{eq:Ws}
\ee

The spin decoherence rates $V_{nm}$ in Eqs.~(\ref{eq:Masterb})
for states $n$ and $m$ with $n,m \in \{1 \dots 4\}$, i.e.
the decoherence rate between states in which both electrons are located in a 
quantum dot, is given by:
\bea
V_{nm} =
\frac{1}{T_{2,D}} + \frac{1}{2} \sum_{j\neq n,m} ( W_{jn} +
W_{jm}) \hspace*{1cm} n,m \in \{1,\dots,4\},
\label{eq:decoherenceD}
\eea
where the W's refer to tunnel rates out of a dot. 
The coherence between state $n$ and $m$ thus not only depends on
the intrinsic spin decoherence time $T_{2,D}$ which is caused by e.g. spin-orbit 
or hyperfine interactions in the dots~\cite{golo04}, but is also reduced by the
(incoherent) tunneling processes from dot to channel~\cite{enge02}. 
Similarly, $V_{nm}$ for all other states $n$ and $m$ is given by
\bea
V_{nm} = \left\{ \ba{l}
\frac{1}{T_{2,C}} + \frac{1}{2} \sum_{j\neq n,m} ( W_{jn} + W_{jm}) 
\hspace*{0.2cm} n,m\in \{1,\dots,16\},\, \mbox{\rm but not both}\ n,m 
\in \{1,\dots,4\}\\
\infty \hspace*{5.cm} n\in \{17,\dots,36\}\ \ \mbox{\rm and/or}\ \ m 
\in \{17,\dots,36\},
\ea \right.
\label{eq:decoherenceC}
\eea
with the W's tunnel rates from a channel to an exit.
Note that energy relaxation processes between different modes in the channels,
i.e. between modes that contribute to the same set of channel states C, do
not affect the transition rates $W_{nm}$ and decoherence rates $V_{nm}$ 
for the states where either $n$ or $m$ or both refer to a channel state.
The reason for this is that these rates refer to resp. {\it spin} flip and 
{\it spin} decoherence processes, which are not affected by {\it orbital} (energy) 
relaxation and decoherence~\cite{tunneling}.
Hence our definition of the channel states as sets of all modes with the 
same spin does not interfere with the definition of spin relaxation and decoherence
of the quantum states.

With the above ingredients, the coupled equations (\ref{eq:Master})
can be solved analytically. We proceed in three steps: ESR applied to the left dot,
 ESR applied to the right dot and the time evolution after
the gates to the quantum channels have been opened. 
 During each step only part of the quantum states are evolving in time, while the 
others remain unchanged. This simplifies the procedure to obtain an analytical solution.

\subsection{Step 1: ESR applied to the left dot}

Initially, at time $t=0$, both spins are assumed 
to be in the singlet state in the quantum dots, so
\bs
\bea
\rho_2(0)& = & \rho_3(0)=1/2;\hspace*{0.5cm} \rho_j(0) = 0 \ \ \ \forall j\in \{1,4,5, \dots, 36 \} \\
\rho_{2,3}(0)& = & \rho_{3,2}(0)=-1/2;\hspace*{0.5cm} \rho_{i,j}(0) = 0 \ \ \ \mbox{\rm otherwise}.
\eea
\label{eq:singlet}
\es
During ESR applied to the left dot quantum states 
$\rho_5(t)-\rho_{36}(t)$ remain unchanged, since the gates between the dots and 
channels are closed. The coherent evolution of $\rho_1(t)$-$\rho_4(t)$ is then 
governed by the Hamiltonian
\bea
{\cal H}_{ESR}(t) = \left( \ba{cccc}
E_1 & \hbar \del_{L} \cos(\omega t) & \hbar \Del_{L}\cos(\omega t) & 0 \nn \\
\hbar \del_{L}\cos(\omega t) & E_2 & 0 & \hbar \Del_{L}\cos(\omega t) \nn \\
\hbar \Del_{L}\cos(\omega t) & 0 & E_2 & \hbar \del_{L}\cos(\omega t) \nn \\
0 & \hbar \Del_{L} \cos(\omega t)  & \hbar \del_{L}\cos(\omega t) & E_4 
\ea \right).
\eea
Including spin-flip rates $W_{D\ua \da}$ and $W_{D\da \ua}$ and the decoherence rate
$\Gam \equiv 1/T_{2,D}$ for both dots~\cite{dec} we then obtain from 
Eqs.~(\ref{eq:Master}) the master equations
\begin{subequations}
\bea
\dot{\rho}_1 & = & - \del_{L} \mbox{\rm Im} \til{\rho}_{1,2} - \Del_{L} \mbox{\rm Im} \til{\rho}_{1,3} - 
2 W_{D\da \ua} \rho_1 + W_{D\ua \da} (\rho_2 + \rho_3) 
\label{eq:masterequations1} 
\\
\dot{\rho}_2 & = & \del_{L} \mbox{\rm Im} \til{\rho}_{1,2} - \Del_{L} \mbox{\rm Im} \til{\rho}_{2,4} + 
W_{D\ua \da} + (W_{D\da \ua}-W_{D\ua \da}) \rho_1 - (2 W_{D\ua \da} + W_{D\da \ua}) 
\rho_2 - W_{D\ua \da} \rho_3 
\\
\dot{\rho}_3 & = & - \del_{L} \mbox{\rm Im} \til{\rho}_{3,4} + \Del_{L} \mbox{\rm Im} \til{\rho}_{1,3} + W_{D\ua \da} + (W_{D\da \ua}-W_{D\ua \da}) \rho_1 - W_{D\ua \da} \rho_2 - (2 W_{D\ua \da} + W_{D\da \ua}) \rho_3 
\\
\mbox{\rm Im} \dot{\til{\rho}}_{1,2} & = & - \frac{\del_{L}}{2}(\rho_2 - \rho_1) + \frac{\Del_{L}}{2} \left( \mbox{\rm Re} \til{\rho}_{1,4} - \mbox{\rm Re} \til{\rho}_{2,3}\right) - \Gam \mbox{\rm Im} \til{\rho}_{1,2}
\label{eq:8d}
\\
\mbox{\rm Im} \dot{\til{\rho}}_{1,3} & = & - \frac{\Del_{L}}{2}(\rho_3 - \rho_1) + \frac{\del_{L}}{2} \left( \mbox{\rm Re} \til{\rho}_{1,4} - \mbox{\rm Re} \til{\rho}_{2,3} \right) - \Gam \mbox{\rm Im} \til{\rho}_{1,3} 
\\
\mbox{\rm Im} \dot{\til{\rho}}_{2,4} & = & - \frac{\Del_{L}}{2}(\rho_4 - \rho_2) - \frac{\del_{L}}{2} \left( \mbox{\rm Re} \til{\rho}_{1,4} - \mbox{\rm Re} \til{\rho}_{2,3} \right) - \Gam \mbox{\rm Im} \til{\rho}_{2,4} 
\\
\mbox{\rm Im} \dot{\til{\rho}}_{3,4} & = & - \frac{\del_{L}}{2}(\rho_4 - \rho_3) - \frac{\Del_{L}}{2} \left( \mbox{\rm Re} \til{\rho}_{1,4} - \mbox{\rm Re} \til{\rho}_{2,3}\right) - \Gam \mbox{\rm Im} \til{\rho}_{3,4} 
\\
\mbox{\rm Re} \dot{\til{\rho}}_{1,4} & = & - \frac{\del_{L}}{2} \left( \mbox{\rm Im} \til{\rho}_{1,3} - \mbox{\rm Im} \til{\rho}_{2,4} \right) - \frac{\Del_{L}}{2} \left( \mbox{\rm Im} \til{\rho}_{1,2} - \mbox{\rm Im} \til{\rho}_{3,4}\right) - \Gam \mbox{\rm Re} \til{\rho}_{1,4} 
\\
\mbox{\rm Re} \dot{\til{\rho}}_{2,3} & = & \frac{\del_{L}}{2} \left( \mbox{\rm Im} \til{\rho}_{1,3} - \mbox{\rm Im} \til{\rho}_{2,4} \right) + \frac{\Del_{L}}{2} \left( \mbox{\rm Im} \til{\rho}_{1,2} - \mbox{\rm Im} \til{\rho}_{3,4} \right) - \Gam \mbox{\rm Re} \til{\rho}_{2,3} 
\label{eq:masterequations9}
\\
\mbox{\rm Re} \dot{\til{\rho}}_{1,2} & = & - \frac{\Del_{L}}{2} \left( \mbox{\rm Im} \til{\rho}_{1,4} + \mbox{\rm Im} \til{\rho}_{2,3} \right) - \Gam \mbox{\rm Re} \til{\rho}_{1,2}
\label{eq:masterequations10}
\\
\mbox{\rm Re} \dot{\til{\rho}}_{1,3} & = & - \frac{\del_{L}}{2} \left( \mbox{\rm Im} 
\til{\rho}_{1,4} - \mbox{\rm Im} \til{\rho}_{2,3}\right) - \Gam \mbox{\rm Re} \til{\rho}_{1,3} 
\\
\mbox{\rm Re} \dot{\til{\rho}}_{2,4} & = & \frac{\del_{L}}{2} \left( \mbox{\rm Im} \til{\rho}_{1,4} - \mbox{\rm Im} \til{\rho}_{2,3} \right) - \Gam \mbox{\rm Re} \til{\rho}_{2,4} 
\\
\mbox{\rm Re} \dot{\til{\rho}}_{3,4} & = & \frac{\Del_{L}}{2} \left( \mbox{\rm Im} \til{\rho}_{1,4} + \mbox{\rm Im} \til{\rho}_{2,3} \right) - \Gam \mbox{\rm Re} \til{\rho}_{3,4} 
\\
\mbox{\rm Im} \dot{\til{\rho}}_{1,4} & = & \frac{\del_{L}}{2} \left( \mbox{\rm Re} \til{\rho}_{1,3} - \mbox{\rm Re} \til{\rho}_{2,4}\right) + \frac{\Del_{L}}{2} \left( \mbox{\rm Re} \til{\rho}_{1,2} - \mbox{\rm Re} \til{\rho}_{3,4}\right) - \Gam \mbox{\rm Im} \til{\rho}_{1,4} 
\\
\mbox{\rm Im} \dot{\til{\rho}}_{2,3} & = & - \frac{\del_{L}}{2} \left( \mbox{\rm Re} \til{\rho}_{1,3} - \mbox{\rm Re} \til{\rho}_{2,4}\right) + \frac{\Del_{L}}{2} \left( \mbox{\rm Re} \til{\rho}_{1,2} - \mbox{\rm Re} \dot{\til{\rho}}_{3,4} \right) - \Gam \mbox{\rm Im} \til{\rho}_{2,3},
\label{eq:masterequations15}
\eea
\label{eq:masterequations}
\end{subequations}
with $\til{\rho}_{i,j}(t) \equiv \rho_{i,j}(t)\, e^{-i\omega t}$ for (ij)\,$\in$\,\{(12), (13), (24), (34)\}, $\til{\rho}_{1,4}(t) \equiv \rho_{1,4}(t)\, e^{-2i\omega t}$ and $\til{\rho}_{2,3}(t) \equiv \rho_{2,3}(t)$.
Eqs.~(\ref{eq:masterequations}) are valid on resonance, so $\hbar \omega \equiv E_2 - E_1 = E_4 - E_2 = \Del E_Z$ and within the rotating wave approximation (RWA)~\cite{RWA}. $\rho_4(t)$ is given by $\rho_4(t) = 1 - \rho_1(t) - \rho_2(t) - \rho_3(t)$.

Eqs.~(\ref{eq:masterequations}) can be split into two sets of coupled equations: Eqs.~(\ref{eq:masterequations1})-(\ref{eq:masterequations9}) and Eqs.~(\ref{eq:masterequations10})-(\ref{eq:masterequations15}). The solution of the second set is straightforwardly obtained and given by
\begin{subequations}
\bea
\mbox{\rm Re} \til{\rho}_{1,2}(t) & = & \frac{1}{2} \left( \mbox{\rm Re} [\til{\rho}_{1,2}(0) - \til{\rho}_{3,4}(0)] \cos (\Del_{L} t) - \mbox{\rm Im} [\til{\rho}_{1,4}(0) + \til{\rho}_{2,3}(0)] \sin (\Del_{L} t) + \right. \nn \\
& & \hspace*{0.5cm} \left. \mbox{\rm Re}[ \til{\rho}_{1,2}(0) + \til{\rho}_{3,4}(0)] \right)\, e^{-\Gam t} 
\\
\mbox{\rm Re} \til{\rho}_{1,3}(t) & = &  \frac{1}{2} \left( \mbox{\rm Re}[ \til{\rho}_{1,3}(0) - \til{\rho}_{2,4}(0)] \cos (\del_{L} t) - \mbox{\rm Im} [\til{\rho}_{1,4}(0) - \til{\rho}_{2,3}(0)] \sin (\del_{L} t) + \right. \nn \\
& & \left. \hspace*{0.5cm} \mbox{\rm Re}[ \til{\rho}_{1,3}(0) + \til{\rho}_{2,4}(0)] \right)\, e^{-\Gam t} 
\\
\mbox{\rm Re} \til{\rho}_{2,4}(t) & = & \frac{1}{2} \left( - \mbox{\rm Re}[ \til{\rho}_{1,3}(0) - \til{\rho}_{2,4}(0)] \cos (\del_{L} t) + \mbox{\rm Im} [\til{\rho}_{1,4}(0) - \til{\rho}_{2,3}(0)] \sin (\del_{L} t) + \right. \nn \\
& & \hspace*{0.5cm} \left. \mbox{\rm Re}[ \til{\rho}_{1,3}(0) + \til{\rho}_{2,4}(0)] \right)\, e^{-\Gam t} 
\\
\mbox{\rm Re} \til{\rho}_{3,4}(t) & = & \frac{1}{2} \left( - \mbox{\rm Re}[ \til{\rho}_{1,2}(0) - \til{\rho}_{3,4}(0)] \cos (\Del_{L} t) + \mbox{\rm Im} [\til{\rho}_{1,4}(0) + \til{\rho}_{2,3}(0)] \sin (\Del_{L} t) + \right. \nn \\ 
& & \hspace*{0.5cm} \left. \mbox{\rm Re}[ \til{\rho}_{1,2}(0) + \til{\rho}_{3,4}(0)] \right)\, 
e^{-\Gam t} 
\\
\mbox{\rm Im} \til{\rho}_{1,4}(t) & = & \frac{1}{2} \left( \mbox{\rm Im} [\til{\rho}_{1,4}(0) + \til{\rho}_{2,3}(0)] \cos (\Del_{L} t) + \mbox{\rm Im} [\til{\rho}_{1,4}(0) - \til{\rho}_{2,3}(0)] \cos (\del_{L} t) + \right. \nn \\
& & \hspace*{0.5cm} \left. \mbox{\rm Re}[ \til{\rho}_{1,2}(0) - \til{\rho}_{3,4}(0)] \sin (\Del_{L} t) + \mbox{\rm Re}[ \til{\rho}_{1,3}(0) - \til{\rho}_{2,4}(0)] \sin(\del_{L}t)  \right)\, 
e^{-\Gam t} 
\\
\mbox{\rm Im} \til{\rho}_{2,3}(t) & = & \frac{1}{2} \left( \mbox{\rm Im} [\til{\rho}_{1,4}(0) + \til{\rho}_{2,3}(0)] \cos (\Del_{L} t) - \mbox{\rm Im} [\til{\rho}_{1,4}(0) - \til{\rho}_{2,3}(0)] \cos (\del_{L} t) + \right. \nn \\
& & \hspace*{0.5cm} \left. \mbox{\rm Re}[ \til{\rho}_{1,2}(0) - \til{\rho}_{3,4}(0)] \sin (\Del_{L} t) - \mbox{\rm Re}[ \til{\rho}_{1,3}(0) - \til{\rho}_{2,4}(0)] \sin(\del_{L}t) \right)\, 
e^{-\Gam t}.
\eea
\label{eq:set8}
\end{subequations}

In order to solve the set of equations (\ref{eq:masterequations1})-(\ref{eq:masterequations9}) we express $\rho_1$-$\rho_3$, $\mbox{\rm Im} \tilde{\rho}_{1,2}$, $\mbox{\rm Im} \tilde{\rho}_{1,3}$, $\mbox{\rm Im} \tilde{\rho}_{2,4}$, $\mbox{\rm Im} \tilde{\rho}_{3,4}$, $\mbox{\rm Re} \tilde{\rho}_{1,4}$ and $\mbox{\rm Re} \tilde{\rho}_{2,3}$ in terms of new variables $x_1$-$x_8$ as follows:
\bea
\rho_1(t) & = & \frac{1}{2} (x_1(t) + x_2(t) - x_3(t))\, e^{-\Gam t} \nn \\
\rho_2(t) & = & \frac{1}{2} (x_1(t) - x_2(t) + x_3(t))\, e^{-\Gam t} 
\nn \\
\rho_3(t) & = & \frac{1}{2} (- x_1(t) + x_2(t) + x_3(t))\, e^{-\Gam t} \nn \\
\mbox{\rm Im} \til{\rho}_{1,2} (t) & = & \frac{1}{2} (x_4(t) + x_6(t))\, e^{-\Gam t} \nn \\
\mbox{\rm Im} \til{\rho}_{1,3} (t) & = & \frac{1}{2} (x_5(t) + x_7(t))\, e^{-\Gam t} 
\label{eq:rhoESR} \\
\mbox{\rm Im} \til{\rho}_{2,4} (t) & = & \frac{1}{2} (x_5(t) - x_7(t))\, e^{-\Gam t}\nn \\
\mbox{\rm Im} \til{\rho}_{3,4} (t) & = & \frac{1}{2} (x_4(t) - x_6(t))\, e^{-\Gam t}\nn \\
\mbox{\rm Re} \til{\rho}_{1,4} (t) & = & x_8(t)\, e^{-\Gam t} \nn \\
\mbox{\rm Re} \til{\rho}_{2,3} (t) & = & (- x_8(t) + Z)\, e^{-\Gam t}, \nn
\eea
with $Z\equiv \mbox{\rm Re} [\tilde{\rho}_{1,4}(0) + \tilde{\rho}_{2,3}(0)]$. The transformation~(\ref{eq:rhoESR}) originates from pairwise adding and subtracting those equations among (\ref{eq:8d})-(\ref{eq:masterequations9}) which share a common term on the right-hand side, e.g. the equations for $\mbox{\rm Im} \dot{\til{\rho}}_{1,2}$ and $\mbox{\rm Im} \dot{\til{\rho}}_{3,4}$. The definition of $x_1$-$x_8$ then naturally arises. Physically, the new variables $x_1$, $x_2$ and $x_3$ can be interpreted as $x_{1(2)}$ = the probability for the spin in the left (right) dot to be up, and $x_3$ = the probability for the two spins to be anti-parallel, each modulated by the exponential dependence on the decoherence rate $\Gamma$. Using (\ref{eq:rhoESR}), Eqs.~(\ref{eq:masterequations1})-(\ref{eq:masterequations9}) are rewritten in terms of $x_1(t)$-$x_8(t)$, which leads to 3 sets of coupled equations. These equations and their solution are given in Appendix~\ref{app-A}. Eqs.~(\ref{eq:rhoESR}) at time $t=t_1$, where $t_1$ is the time during which the ESR field is switched on, thus represent the density matrix elements for the double dot states after the ESR rotation applied to the left dot. 

\subsection{Step 2: ESR applied to the right dot}

\begin{figure}
\centerline{\epsfig{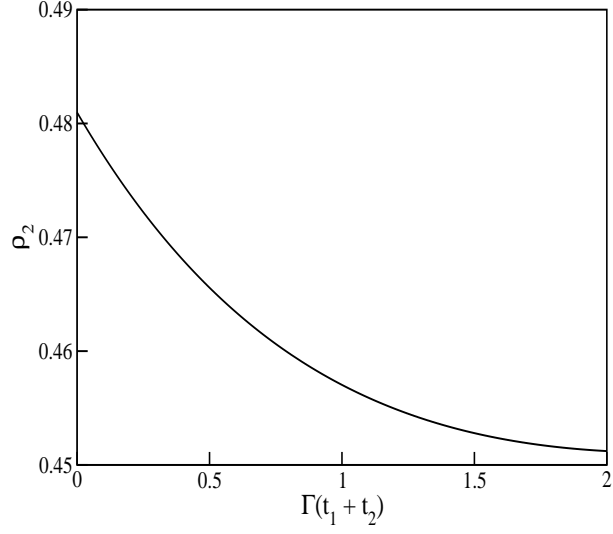}}
\caption[]{The occupation probability $\rho_2(t_1+t_2)$ [Eq.~(\ref{eq:rho2})] of the quantum state D$\ua$D$\da$ as
a function of the amount of decoherence [in units of $1/(t_1 + t_2)]$. For $\Gam$=0 $\rho_2$ is given by Eq.~(\ref{eq:rho2special}). Parameters used are
$\Del\! \cdot\! t_1 = \pi/4$, $\Del\! \cdot\! t_2=\pi/8$ and all W's=0.} 
\label{fig:plotrho}
\end{figure}

Eqs.~(\ref{eq:rhoESR}) can also directly be used to obtain the solution after the second ESR rotation applied to the right dot, by substituting $\Del_L \rightarrow \del_R$ and $\del_L \rightarrow \Del_R$ in Eqs.~(\ref{eq:set8}) and (\ref{eq:setx1}), and by exchanging $x_6 \leftrightarrow x_7$ in Eqs.~(\ref{eq:setx2}), using $\rho_1(t_1)$ instead of $\rho_1(0)$ etc. as initial conditions. In order to illustrate this solution, let us consider the initial condition of a singlet in the double dot [Eq.~(\ref{eq:singlet})] and let $t_2$ be the duration of the second ESR rotation. In case of no dissipation (all $W$'s = 0) and no influence of ESR applied to one dot on the spin in the other dot we then obtain from Eqs.~(\ref{eq:rhoESR}), (\ref{eq:setx1}) and (\ref{eq:setx2}) for e.g. the occupation probability $\rho_2(t_1+t_2)$ the expression:
\bea
\rho_2(t_1+t_2) & = & \frac{1}{4}\, \left[ 1 +  \left\{ [\cos 
(\tilde{\Omega}_{\Del} t_1) + \frac{\Gam}{2 \tilde{\Omega}_{\Del}} 
\sin(\tilde{\Omega}_{\Del} t_1)]\, [\cos(\tilde{\Omega}_{\Del} t_2) + 
\frac{\Gam}{2 \tilde{\Omega}_{\Del}} \sin(\tilde{\Omega}_{\Del} t_2) ] +
\right. \right. \nn \\
& & \ \ \ \left. \left. \frac{\Del}{\tilde{\Omega}_{\Del}} \sin(\Del\! \cdot\! t_1) 
\sin(\tilde{\Omega}_{\Del} t_2)  
e^{- \frac{\Gam}{2} t_1} \right\} e^{-\frac{\Gam}{2} (t_1+t_2)} \right]
\label{eq:rho2}
\eea
with $\tilde{\Omega}_{\Del} \equiv \frac{1}{2} \sqrt{4\Del^2 - \Gam^2}$ and $\Del \equiv \Del_L=\Del_R$. In the absence of decoherence ($\Gam  = 0$) the expressions for $\rho_2(t_1+t_2)$ and the other density matrix elements simplify to:
\bs
\bea
\rho_1(t_1 + t_2) & = & \rho_4(t_1 + t_2) = \frac{1}{4} (1 - \cos \theta_1 \cos \theta_2 - \sin \theta_1 \sin \theta_2) \\
\rho_2(t_1 + t_2) & = & \rho_3(t_1 + t_2) = \frac{1}{4} (1 + \cos \theta_1 \cos \theta_2 + \sin \theta_1 \sin \theta_2) 
\label{eq:rho2special} \\
\rho_{1,2} (t_1 + t_2) & = & \rho_{2,4} (t_1 + t_2) = - \frac{i}{4} (\cos \theta_1 \sin \theta_2 - \sin \theta_1 \cos \theta_2) \\
\rho_{1,3} (t_1 + t_2) & = & \rho_{3,4} (t_1 + t_2) = \frac{i}{4} (\cos \theta_1 \sin \theta_2 - \sin \theta_1 \cos \theta_2) \\
\rho_{1,4}(t_1 + t_2) & = & - \frac{1}{4} (1 - \cos \theta_1 \cos \theta_2 - \sin \theta_1 \sin \theta_2) \\
\rho_{2,3}(t_1 + t_2) & = & - \frac{1}{4} (1 + \cos \theta_1 \cos \theta_2 + \sin \theta_1 \sin \theta_2),
\eea
\es
with $\theta_1 \equiv \tilde{\Omega}_{\Del} t_1$ and $\theta_2 \equiv \tilde{\Omega}_{\Del} 
t_2$. Eq.~(\ref{eq:rho2}) is plotted in Fig.~\ref{fig:plotrho} as a function of the amount of decoherence $\Gam$.

Already for moderate amounts of decoherence $\Gam (t_1 + t_2)=0.001$ the occupation probability
has become 0.01\% less than its value in the absence of decoherence $\rho_2^{\Gam=0}
(t_1+t_2)=0.481$ for the set of parametes chosen in Fig.~\ref{fig:plotrho}. 
This increases to 0.1\% for $\Gam (t_1 + t_2)=0.01$. 

\subsection{Step 3: Time evolution after the gates to the channels have been opened}

We now turn to the next step in the evolution of the entangled pair in Fig.~\ref{fig:system}, namely the time evolution of the density matrix elements after the ESR rotations are completed and the gates to the quantum channels are opened, see Fig.~\ref{fig:system}(c). From this moment onwards the coherent evolution due to the first term on the RHS of Eqs.~(\ref{eq:Master}) stops and the time evolution of the matrix elements is solely determined by decay and decoherence rates represented by the second terms on the RHS of Eqs.~(\ref{eq:Master}). The off-diagonal elements $\rho_{i,j}(t)$ then rotate with $(E_i - E_j)/\hbar$ and decay with rate $V_{ij}$:
\be
\rho_{i,j}(t) = \rho_{i,j}(t_{ESR})\, e^{i(E_i - E_j)(t - t_{ESR})/\hbar}\, e^{-V_{ij}(t - t_{ESR})} \hspace*{1cm} \mbox{for}\ t \geq t_{ESR},
\label{eq:coh}
\ee
where $t_{ESR} \equiv t_1 + t_2$ and $V_{ij}$ is given by Eq.~(\ref{eq:decoherenceD}) for 
$i,j \in \{1\dots 4\}$ and Eq.~(\ref{eq:decoherenceC}) otherwise. The initial values $\rho_{i,j}(t_{ESR})$ for $i,j \in \{1, \dots, 16\}$ are given by $\rho_{m,n}(t_{ESR})$ for 
$m,n \in \{1\dots 4\}$ [Eqs.~(\ref{eq:rhoESR})] with the correspondence in indices
\bea 
i(j) & \in & \{1,5,9,13\} \leftrightarrow m(n)=1 \nn \\
i(j) & \in & \{2,6,10,14\} \leftrightarrow m(n)=2 \nn \\
i(j) & \in & \{3,7,11,15\} \leftrightarrow m(n)=3 \nn \\
i(j) & \in & \{4,8,12,16\} \leftrightarrow m(n)=4. \nn
\eea
In this way the coherence at time $t_{ESR}$ between any pair of states $i,j \in \{1, \dots, 16\}$ is given by the coherence at $t_{ESR}$ between those dot states m,n $\in \{1, \dots, 4\}$ which can (eventually) coherently evolve into i and j, i.e. the dot states m and n which have the same spin states as i and j respectively. So, for example, $\rho_{C\ua C\ua, D\da C\da} (t_{ESR}) = 
\rho_{D\ua D\ua, D\da D\da} (t_{ESR}) \equiv \rho_{1,4}(t_{ESR})$. Note that $\rho_{i,j}(t)=0$ for those states in which at least one electron has reached a detector ($i \in \{17, \dots, 36\}$ and/or $j \in \{17, \dots, 36\}$), since for those states $V_{ij}=\infty$. This corresponds to the assumption of immediate detection.

In the remaining part of this paper we focus on the evolution of the populations $\rho_1(t)$-$\rho_{36}(t)$ for times $t\geq t_{ESR}$ under the following conditions:
\begin{itemize}
\item{We neglect the possibility of spin flips in the dots, i.e. we set $W_{D\ua \da}=W_{D\da \ua}=0$. This based on the fact that $T_{1,D}$ is known to be much longer (0.85 ms at magnetic fields $B_z=8 T$)~\cite{elze04} than the time required to travel through the channels to the exits. This assumption is not essential to obtain an analytical solution; it only simplifies the resulting equations.
}
\item{We assume that the tunnel rate $W_T$ out of the dots into the channels is equal for spin-up and spin-down electrons, i.e. the two electrons tunnel out of the singlet state with a negligible time delay $t_{delay}$ in between, and that spin is conserved during this tunneling process. Typically~\cite{cerl04} $t_{delay} \approx 10^{-13}$ s, which is much less than the travel time through a channel $\sim 10^{-10} s$.
}
\item{The tunnel rate $W_E$ through the QPCs is taken to be constant and equal for spin-up and spin-down electrons, i.e. the set-up is assumed to be constructed in such a way that the detection time for spin-up and spin-down electrons once they have reached the QPCs is the same.
}
\item{Spin flips in the exits are neglected, i.e. detection is assumed to be faster (with typical times $\sim 10^{-11}$\, s)~\cite{debl03} than the spin-flip rate ($\gg 10^{-11}$\, s)~\cite{blaa05} in the detectors.
}
\end{itemize}

The evolution equations for $\rho_1(t)$-$\rho_{36}(t)$ for times $t\geq t_{ESR}$ are then given by the master equations
\be
\dot{\rho}_i  =  -2 W_T \rho_i \ \ \mbox{\rm for}\ \ i \in\{1, \dots, 4\}
\label{eq:1to4}
\ee
\bs
\bea
\dot{\rho}_i & = & W_T \rho_{i-4} + W_{C\ua \da}\rho_{i+2} - (W_E + W_T + W_{C\da \ua}) \rho_i \hspace*{1cm} i\in \{5,6 \} \\
\dot{\rho}_i & = & W_T \rho_{i-4} + W_{C\da \ua}\rho_{i-2} - (W_E + W_T + W_{C\ua \da}) \rho_i \hspace*{1cm} i\in \{7,8 \} \\
\dot{\rho}_i & = & W_T \rho_{i-8} + W_{C\ua \da}\rho_{i+1} - (W_E + W_T + W_{C\da \ua}) \rho_i \hspace*{1cm} i\in \{9,11 \} \\
\dot{\rho}_i & = & W_T \rho_{i-8} + W_{C\da \ua}\rho_{i-1} - (W_E + W_T + W_{C\ua \da}) \rho_i \hspace*{1cm} i\in \{10,12 \} 
\eea
\label{eq:5to12}
\es
\bs
\bea
\dot{\rho}_{13} & = & W_T (\rho_5 + \rho_9) + W_{C\ua \da} (\rho_{14} + \rho_{15}) 
- 2 (W_E + W_{C\da \ua}) \rho_{13} \\ 
\dot{\rho}_{14} & = & W_T (\rho_6 + \rho_{10}) + W_{C\da \ua}\rho_{13} + W_{C\ua \da}\rho_{16} 
- (2 W_E + W_{C\ua \da} + W_{C\da \ua}) \rho_{14} \\ 
\dot{\rho}_{15} & = & W_T (\rho_7 + \rho_{11}) + W_{C\da \ua}\rho_{13} + W_{C\ua \da}\rho_{16} 
- (2 W_E + W_{C\ua \da} + W_{C\da \ua}) \rho_{15} \\
\dot{\rho}_{16} & = & W_T (\rho_8 + \rho_{12}) + W_{C\da \ua} (\rho_{14} + \rho_{15}) 
- 2 (W_E + W_{C\ua \da}) \rho_{16} 
\eea
\label{eq:13to16}
\es
\be
\dot{\rho}_{i}(t) = W_E \rho_{i-12}(t - t_{travel} + t_{ESR}) - W_T \rho_i(t)
\hspace*{1cm} i\in \{17, \dots, 24 \}
\label{eq:17to24}
\ee
\bs
\bea
\dot{\rho}_i(t) & = & W_T \rho_{i-8}(t) + W_E \rho_{i-12}(t - t_{travel} + t_{ESR}) + W_{C\ua \da}\rho_{i+1}(t) - (W_E + W_{C\da \ua}) \rho_i(t) \nn \\
& & \mbox{\rm for}\ i\in \{25,27 \} \\
\dot{\rho}_i(t) & = & W_T \rho_{i-8}(t) + W_E \rho_{i-12}(t - t_{travel} + t_{ESR}) + W_{C\da \ua}\rho_{i-1}(t) - (W_E + W_{C\ua \da}) \rho_i(t) \nn \\
& & \mbox{\rm for}\  i\in \{26,28 \} \\ 
\dot{\rho}_i(t) & = & W_T \rho_{i-8}(t) + W_E \rho_{i-16}(t - t_{travel} + t_{ESR}) + W_{C\ua \da}\rho_{i+2}(t) - (W_E + W_{C\da \ua}) \rho_i(t) \nn \\
& & \mbox{\rm for}\ i\in \{29,30 \} \\
\dot{\rho}_i(t) & = & W_T \rho_{i-8}(t) + W_E \rho_{i-16}(t - t_{travel} + t_{ESR}) + W_{C\da \ua}\rho_{i-2}(t) - (W_E + W_{C\ua \da}) \rho_i(t) \nn \\
& & \mbox{\rm for}\ i\in \{31,32 \}
\eea
\label{eq:25to32}
\es
\be
\dot{\rho}_i  =  W_E (\rho_{i-8} + \rho_{i-4}) \hspace*{1cm} i\in \{33, \dots, 36 \}.
\label{eq:33to36}
\ee
Here $t_{travel}> t_{ESR}$ denotes the earliest time at which an electron has traveled through the channels and reached an exit. For times $t\leq t_{travel}$, $\rho_1(t)-\rho_{36}(t)$ is thus given by Eqs.~(\ref{eq:1to4})-(\ref{eq:33to36}) for $W_E=0$, since at those times no electron can have arrived at a detector yet. The above sets of coupled equations can be solved one by one: first those for $\rho_1(t)$-$\rho_4(t)$, then once the latter are known those for $\rho_5(t)$-$\rho_{12}(t)$ (in the pairs (5,7), (6,8), (9,10) and (11,12)), then $\rho_{13}(t)$-$\rho_{16}(t)$ and $\rho_{17}(t)$-$\rho_{24}(t)$, subsequently $\rho_{25}(t)$-$\rho_{32}(t)$ (in the pairs (25,26), (27,28), (29,31) and (30,32)) and finally $\rho_{33}(t)$-$\rho_{36}(t)$. Proceeding in this order and using initial conditions
\be
\rho_i(t_{ESR}) =  \left\{ \ba{ll} 
\mbox{\rm Eqs.}~(\ref{eq:rhoESR}) \ \ \mbox{for}\ i=1, \dots, 4 \\
0 \hspace*{1.8cm} \mbox{for}\ i=5, \dots, 36 
\ea
\right.
\ee
we obtain for $\rho_1(t)$-$\rho_4(t)$, the states in which both electrons are located in a dot:
\be
\rho_i(t) =  \rho_i(t_{ESR})\, e^{-2 W_T (t - t_{ESR})} \hspace*{0.7cm} i \in\{1, \dots, 4\},\ 
t\geq t_{ESR}.
\label{eq:sol1to4}
\ee
Next, we find for $\rho_5(t)$-$\rho_{12}(t)$, which correspond to the quantum states in which one electron is located in a dot and the other in a channel, from Eqs.~(\ref{eq:5to12}):
\bs
\bea
\rho_5(t) & = & A_{5,7,1,3} e^{-W_{ETC}(t-t_{ESR})} +
B_{5,7,1,3}\, e^{-(W_E + W_T)(t-t_{ESR})} + C_{1,3}\, e^{- 2 W_T (t-t_{ESR})} \\
\rho_6(t) & = & A_{6,8,2,4} e^{-W_{ETC}(t-t_{ESR})} +
B_{6,8,2,4}\, e^{-(W_E + W_T)(t-t_{ESR})} + C_{2,4}\, e^{- 2 W_T (t-t_{ESR})} \\
\rho_7(t) & = & - A_{5,7,1,3} e^{-W_{ETC}(t-t_{ESR})} +
\frac{W_{C\da\ua}}{W_{C\ua\da}} B_{5,7,1,3}\, e^{-(W_E + W_T)(t-t_{ESR})} + 
D_{1,3}\, e^{- 2 W_T (t-t_{ESR})} \\
\rho_8(t) & = & - A_{6,8,2,4} e^{-W_{ETC}(t-t_{ESR})} +
\frac{W_{C\da\ua}}{W_{C\ua\da}} B_{6,8,2,4}\, e^{-(W_E + W_T)(t-t_{ESR})} + 
D_{2,4}\, e^{- 2 W_T (t-t_{ESR})} \\
\rho_9(t) & = & A_{9,10,1,2} e^{-W_{ETC}(t-t_{ESR})} +
B_{9,10,1,2}\, e^{-(W_E + W_T)(t-t_{ESR})} + C_{1,2}\, e^{- 2 W_T (t-t_{ESR})} \\
\rho_{10}(t) & = & - A_{9,10,1,2} e^{-W_{ETC}(t-t_{ESR})} +
\frac{W_{C\da\ua}}{W_{C\ua\da}} B_{9,10,1,2}\, e^{-(W_E + W_T)(t-t_{ESR})} + 
D_{1,2}\, e^{- 2 W_T (t-t_{ESR})} \\
\rho_{11}(t) & = & A_{11,12,3,4} e^{-W_{ETC}(t-t_{ESR})} +
B_{11,12,3,4}\, e^{-(W_E + W_T)(t-t_{ESR})} + C_{3,4}\, e^{- 2 W_T (t-t_{ESR})} \\
\rho_{12}(t) & = & - A_{11,12,3,4} e^{-W_{ETC}(t-t_{ESR})} +
\frac{W_{C\da\ua}}{W_{C\ua\da}} B_{11,12,3,4}\, e^{-(W_E + W_T)(t-t_{ESR})} + 
D_{3,4}\, e^{- 2 W_T (t-t_{ESR})},
\eea
\label{eq:sol5to12}
\es
where
\bs
\bea
W_{ETC} & \equiv & W_E + W_T + W_{C\ua \da} + W_{C\da \ua} \\
A_{i,j,k,l} & \equiv & \frac{W_{C\da\ua}\rho_i(t_{ESR}) - W_{C\ua\da}\rho_j(t_{ESR})}{W_{C\ua\da} + W_{C\da\ua}} +
\frac{W_T(- W_{C\da\ua}\rho_k(t_{ESR}) + W_{C\ua\da}\rho_l(t_{ESR}))}{(W_{C\ua\da} + W_{C\da\ua})(W_E - W_T + W_{C\ua\da} + W_{C\da\ua})} \\
B_{i,j,k,l} & \equiv & \frac{W_{C\ua\da}}{W_{C\ua\da} + W_{C\da\ua}} [\rho_i(t_{ESR}) +
\rho_j(t_{ESR}) - \frac{W_T}{W_E - W_T} (\rho_k(t_{ESR}) + \rho_l(t_{ESR}))] \\
C_{i,j} & = & \frac{W_T}{W_E - W_T}\, \frac{(W_E - W_T + W_{C\ua\da}) \rho_i(t_{ESR}) +
W_{C\ua\da} \rho_j(t_{ESR})}{W_E - W_T + W_{C\ua\da} + W_{C\da\ua}} \\
D_{i,j} & = & \frac{W_T}{W_E - W_T}\, \frac{W_{C\da\ua} \rho_i(t_{ESR}) + (W_E - W_T + W_{C\da\ua}) \rho_j(t_{ESR})}{W_E - W_T + W_{C\ua\da} + W_{C\da\ua}}.
\eea
\label{eq:coefAtoD}
\es
For times $t\leq t_{travel}$, the evolution of $\rho_5(t)$-$\rho_{12}(t)$ are given by Eqs.~(\ref{eq:sol5to12}) with $W_E=0$. For times $t\geq t_{travel}$ these populations are given by Eqs.~(\ref{eq:sol5to12}) with $t_{ESR} \rightarrow t_{travel}$. 

In order to obtain the solution for $\rho_{13}(t)$-$\rho_{16}(t)$, which corrresponds to the situation in which both electrons are located in a channel, we rewrite the equations for 
$\dot{\rho}_{13}$-$\dot{\rho}_{16}$ as
\bs
\bea
\dot{\rho}_{13} & = & W_T (\rho_5 + \rho_9) + W_{C\ua \da}(\rho_{14}+\rho_{15}) - 2(W_E + W_{C\da \ua}) \rho_{13} 
\label{eq:13to16a} \\ 
\dot{\rho}_{14}+ \dot{\rho}_{15} & = & W_T (\rho_6 + \rho_7 + \rho_{10} + \rho_{11}) + 
2 W_{C\da \ua}\rho_{13} - (2 W_E + W_{C\ua \da} + W_{C\da \ua}) (\rho_{14} + \rho_{15}) + 2 W_{C\ua \da}\rho_{16} \\
\dot{\rho}_{16} & = & W_T (\rho_8 + \rho_{12}) + W_{C\da \ua}(\rho_{14}+\rho_{15}) - 2(W_E + W_{C\ua \da}) \rho_{16} 
\label{eq:13to16c}\\ 
\dot{\rho}_{14}- \dot{\rho}_{15} & = & W_T (\rho_6 - \rho_{7}  + \rho_{10} - \rho_{11}) - (2 W_E + W_{C\ua \da} + W_{C\da \ua}) (\rho_{14} - \rho_{15}).
\label{eq:13to16d}
\eea
\label{eq:13to16re}
\es
Eqs.~(\ref{eq:13to16re}) consist of 3 coupled equations~(\ref{eq:13to16a})-(\ref{eq:13to16c}) and a separate one, Eq.~(\ref{eq:13to16d}). We first solve the latter and then the first three. In each case the solution is a combination of a homogeneous and a particular solution. Taking from now onwards $W_{C\ua \da}$=$W_{C\da \ua} \equiv W_C$~\cite{blaa052} we obtain:
\bs
\bea
\rho_{13}(t) & = & -E\, e^{-2(W_E+W_C)(t-t_{ESR})} + \frac{1}{2}F\, e^{-2 W_E (t-t_{ESR})} - \frac{1}{2} \tilde{F}\, e^{-2(W_E+2W_C)(t-t_{ESR})} \nn \\ 
& + & H_{13}\, e^{-(W_E+W_T+2W_C)(t-t_{ESR})}
+ K_{13}\, e^{-(W_E+W_T)(t-t_{ESR})} + L_{13}\, e^{-2 W_T (t-t_{ESR})} \\
\rho_{14}(t) & = & \tilde{E}\, e^{-2(W_E+W_C)(t-t_{ESR})} + \frac{1}{2}F\, e^{-2 W_E (t-t_{ESR})} + \frac{1}{2} \tilde{F}\, e^{-2(W_E+2W_C)(t-t_{ESR})} \nn \\
& + & H_{14}\, e^{-(W_E+W_T+2W_C)(t-t_{ESR})} + K_{14}\, e^{-(W_E+W_T)(t-t_{ESR})} + L_{14}\, e^{-2 W_T (t-t_{ESR})} \\
\rho_{15}(t) & = & - \tilde{E}\, e^{-2(W_E+W_C)(t-t_{ESR})} + \frac{1}{2}F\, e^{-2 W_E (t-t_{ESR})} + \frac{1}{2} \tilde{F}\, e^{-2(W_E+2W_C)(t-t_{ESR})} \nn \\ 
& + & H_{15}\, e^{-(W_E+W_T+2W_C)(t-t_{ESR})} + K_{15}\, e^{-(W_E+W_T)(t-t_{ESR})} + L_{15}\, e^{-2 W_T (t-t_{ESR})} \\
\rho_{16}(t) & = & E\, e^{-2(W_E+W_C)(t-t_{ESR})} + \frac{1}{2}F\, e^{-2 W_E (t-t_{ESR})} - \frac{1}{2} \tilde{F}\, e^{-2(W_E+2W_C)(t-t_{ESR})} \nn \\
& + &  H_{16}\, e^{-(W_E+W_T+2W_C)(t-t_{ESR})} + K_{16}\, e^{-(W_E+W_T)(t-t_{ESR})} + L_{16}\, e^{-2 W_T (t-t_{ESR})}.
\eea
\label{eq:sol13to16}
\es
The coefficients in Eqs.~(\ref{eq:sol13to16}) are given in Appendix~\ref{app-B}. Also here, $\rho_{13}(t)$-$\rho_{16}(t)$ for times $t\leq t_{travel}$ are given by Eqs.~(\ref{eq:sol13to16}) with $W_E=0$, and for times $t\geq t_{travel}$ these populations are given by Eqs.~(\ref{eq:sol13to16}) with $t_{ESR} \rightarrow t_{travel}$.

The solution of the next set, $\rho_{17}(t)-\rho_{24}(t)$, corresponding to the states in which one electron is located in a dot while the other has reached a detector, is given by
\bea
\rho_{i}(t) & = & A_i\, e^{-(W_E+W_T+2W_C)(t-t_{travel})} - B_i\, e^{-(W_E+W_T)(t-t_{travel})} - C_i\, e^{-2 W_T(t-t_{travel})} + \nn \\
& & [\rho_i(t_{travel}) - A_i + B_i + C_i]\, e^{- W_T(t-t_{travel})} \ \ \  \mbox{\rm for}\ i\in \{17, \dots, 24\},\ \, t\geq t_{travel}
\label{eq:sol17to24}
\eea
and $\rho_i(t)=0$ for $t\leq t_{travel}$. The coefficients $A_i$, $B_i$ and $C_i$ in Eqs.~(\ref{eq:sol17to24}) are given in Table~\ref{table:coefficients}.
\begin{table}
\begin{tabular}{|c|c|c|c|}
\hline \hline
$i$ & $A_i$ & $B_i$ & $C_i$ \\ \hline \hline
17 & $\frac{-W_E A_{5,7,1,3}}{W_E + 2 W_C}$ & $B_{5,7,1,3}$ & $\frac{W_E}{W_T} C_{1,3}$
 \\ \hline
18 & $\frac{-W_E A_{6,8,2,4}}{W_E + 2 W_C}$ & $B_{6,8,2,4}$ & $\frac{W_E}{W_T} C_{2,4}$
\\ \hline
19 & $\frac{W_E A_{5,7,1,3}}{W_E + 2 W_C}$ & $B_{5,7,1,3}$ & $\frac{W_E}{W_T} D_{1,3}$
 \\ \hline
20 & $\frac{W_E A_{6,8,2,4}}{W_E + 2 W_C}$ & $B_{6,8,2,4}$ & $\frac{W_E}{W_T} D_{2,4}$
\\ \hline
21 & $\frac{-W_E A_{9,10,1,2}}{W_E + 2 W_C}$ & $B_{9,10,1,2}$ & $\frac{W_E}{W_T} C_{1,2}$
\\ \hline
22 & $\frac{W_E A_{9,10,1,2}}{W_E + 2 W_C}$ & $B_{9,10,1,2}$ & $\frac{W_E}{W_T} D_{1,2}$
\\ \hline
23 & $\frac{- W_E A_{11,12,3,4}}{W_E + 2 W_C}$ & $B_{11,12,3,4}$ & $\frac{W_E}{W_T} 
C_{3,4}$
\\ \hline
24 & $\frac{W_E A_{11,12,3,4}}{W_E + 2 W_C}$ & $B_{11,12,3,4}$ & $\frac{W_E}{W_T} 
D_{3,4}$
\\ \hline \hline
\end{tabular}
\caption{Coefficients $A_i$, $B_i$ and $C_i$ in Eqs.~(\ref{eq:sol17to24})}
\label{table:coefficients}
\end{table}
Next, we solve for $\rho_{25}(t)$-$\rho_{32}(t)$, the states in which one spin has reached a detector, while the other is still in a channel, in the pairs $\rho_i(t)$\&$\rho_j(t)$ $\in$ 
$\{ \rho_{25}(t)$\&$\rho_{26}(t)$, $\rho_{27}(t)$\&$\rho_{28}(t)$,  
$\rho_{29}(t)$\&$\rho_{31}(t)$, and  $\rho_{30}(t)$\&$\rho_{32}(t)\}$, see 
Eqs.~(\ref{eq:25to32}). For each pair the solution is given by, for times 
$t \geq t_{travel}$:
\bs
\bea
\rho_{i}(t) & = & P_{i,j}\, e^{-W_E (t-t_{travel})} + Q_{i,j}\, e^{-(W_E+2W_C)(t-t_{travel})} 
+ M_{i,1}\, e^{-2(W_E+W_C)(t-t_{travel})} + \nn \\
& & M_{i,2}\, e^{-2 W_E (t-t_{travel})} + M_{i,3}\, e^{-2(W_E+2W_C)(t-t_{travel})} 
+ M_{i,4}\, e^{-(W_E+W_T+2W_C)(t-t_{travel})} + \nn \\
& & M_{i,5}\, e^{-(W_E+W_T)(t-t_{travel})} + M_{i,6}\, e^{-2 W_T (t-t_{travel})}
+ M_{i,7}\, e^{- W_T (t-t_{travel})} \\
\rho_{j}(t) & = & P_{j,i}\, e^{-W_E (t-t_{travel})} + Q_{j,i}\, e^{-(W_E+2W_C)(t-t_{travel})} 
+ M_{j,1}\, e^{-2(W_E+W_C)(t-t_{travel})} + \nn \\
& & M_{j,2}\, e^{-2 W_E (t-t_{travel})} + M_{j,3}\, e^{-2(W_E+2W_C)(t-t_{travel})} 
+ M_{j,4}\, e^{-(W_E+W_T+2W_C)(t-t_{travel})} + \nn \\
& & M_{j,5}\, e^{-(W_E+W_T)(t-t_{travel})} + M_{j,6}\, e^{-2 W_T (t-t_{travel})}
+ M_{j,7}\, e^{- W_T (t-t_{travel})},
\eea
\label{eq:sol25to32}
\es
and $\rho_i(t) = \rho_j(t)=0$ for $t\leq t_{travel}$. The coefficients $P_{i,j}$, $Q_{i,j}$ and $M_{i,1} \dots M_{i,7}$ for $i,j \in \{25, \dots, 32 \}$ are given in Appendix~\ref{app-B}.

Finally, we obtain the time evolution of the states $\rho_{33}(t)-\rho_{36}(t)$ in which both electrons have reached an exit. This is given by, for times $t\geq t_{travel}$,
\bea
\rho_j(t) & = & - W_E \left\{ \frac{P_{m,p} + P_{n,q}}{W_E}\, e^{-W_E (t-t_{travel})} + 
\frac{Q_{m,p} + Q_{n,q}}{W_E + 2 W_C}\, e^{-(W_E+2W_C) (t-t_{travel})} \right. \nn \\
& + & \frac{M_{m,1} + M_{n,1}}{2(W_E + W_C)}\, e^{-2(W_E+W_C) (t-t_{travel})} +
 \frac{M_{m,2} + M_{n,2}}{2W_E}\, e^{-2 W_E (t-t_{travel})} \nn \\
& + & \frac{M_{m,3} + M_{n,3}}{2(W_E + 2 W_C)}\, e^{-2(W_E+2W_C) (t-t_{travel})} +
 \frac{M_{m,4} + M_{n,4}}{W_E + W_T + 2 W_C}\, e^{-(W_E+W_T+2W_C) (t-t_{travel})} \nn \\
& + & \frac{M_{m,5} + M_{n,5}}{W_E + W_T}\, e^{-(W_E+W_T) (t-t_{travel})} +
 \frac{M_{m,6} + M_{n,6}}{2 W_T}\, e^{-2W_T (t-t_{travel})} \nn \\
& + & \left. \frac{M_{m,7} + M_{n,7}}{W_T}\, e^{-W_T (t-t_{travel})} \right\} 
 \nn \\
& + & W_E(\ \mbox{\rm sum of all previous coefficients, so}\ \frac{P_{m,p} + P_{n,q}}{W_E}
+ \frac{Q_{m,p} + Q_{n,q}}{W_E + 2 W_C} + \dots ) 
\label{eq:sol33to36}
\eea 
for
\bd
(j,m,n,p,q) \in \{ (33,25,29,26,31), (34,26,30,25,32), (35,27,31,28,29), 
(36,28,32,27,30) \}.
\ed
{\it Special case.} In order to illustrate the solution (\ref{eq:sol33to36}), we now derive explicit expressions for $\rho_{33}(t)$ and $\rho_{34}(t)$, the probabilities that a spin-up is detected in the left detector and resp. a spin-up or a spin-down in the right detector, for the special case of $\Gam=0$ (no decoherence in the dots) and $W_{C\ua \da} = W_{C\da \ua}=0$ (no relaxation in the channel). This corresponds to the situation in which the time evolution occurs in the absence of any decoherence and dissipation mechanisms in the dots and channels and only depends on $W_T$, the tunnel rate from dot to channel, and $W_E$, the tunnel rate from channel to exit.

We are interested in finding $\rho_{33}(t)$ and $\rho_{34}(t)$ for times $t$ $\geq$ $t_{travel}$ [since $\rho_{33}(t)$=$\rho_{34}(t)$=0 $\forall$ $t \leq t_{travel}$]. To that end, we first calculate $\rho_j(t_{travel})$ for $j\leq 16$ from Eqs.~(\ref{eq:sol1to4}), 
(\ref{eq:sol5to12}) and (\ref{eq:sol13to16}) and then all coefficients entering the expressions for $\rho_{33}(t)$ and $\rho_{34}(t)$ in Eqs.~(\ref{eq:sol33to36}).

For $\rho_1(t_{travel})$-$\rho_{16}(t_{travel})$ we then obtain:
\bs 
\bea
\rho_1(t_{travel}) & = & \rho_4(t_{travel}) = \rho_1(t_{ESR})\, 
e^{-2W_T(t_{travel}-t_{ESR})} \\
\rho_2(t_{travel}) & = & \rho_3(t_{travel}) = \rho_2(t_{ESR})\, 
e^{-2W_T(t_{travel}-t_{ESR})} \\
\rho_i(t_{travel}) & = & \rho_1(t_{ESR})\, e^{-W_T(t_{travel}-t_{ESR})} (1 - e^{-W_T(t_{travel}-t_{ESR})}) \hspace*{0.4cm}  i\in \{5,8,9,12\} \\
\rho_i(t_{travel}) & = & \rho_2(t_{ESR})\, e^{-W_T(t_{travel}-t_{ESR})} (1 - e^{-W_T(t_{travel}-t_{ESR})}) \hspace*{0.4cm}  i\in \{6,7,10,11\} \\
\rho_{13}(t_{travel}) & = & \rho_{16}(t_{travel}) = \rho_1(t_{ESR})\, 
(1-e^{-W_T(t_{travel}-t_{ESR})})^2 \\
\rho_{14}(t_{travel}) & = & \rho_{15}(t_{travel}) = \rho_2(t_{ESR})\, 
(1-e^{-W_T(t_{travel}-t_{ESR})})^2. 
\eea
\label{eq:initial}
\es
Eqs.~(\ref{eq:initial}) form the initial conditions that appear in the expressions for 
$\rho_{33}(t)$-$\rho_{36}(t)$ [Eqs.~(\ref{eq:sol33to36})]. We then
find $\forall$ $t \geq  t_{travel}$:
\bs
\bea
\rho_{33}(t) & = & 
\left( \rho_{13}(t_{travel}) - \frac{2W_T}{W_E - W_T}\, \rho_5(t_{travel})
+ \frac{W^2_T}{(W_E - W_T)^2}\, \rho_1(t_{travel}) \right)\, e^{-2W_E(t-t_{travel})} + 
\nn \\ 
& & \left( -2\, \rho_{13}(t_{travel})  - \frac{2(W_E - 2W_T)}{W_E - W_T}\, 
\rho_5(t_{travel}) + \frac{2 W_T}{W_E - W_T}\, \rho_1(t_{travel}) \right)\, 
e^{-W_E(t-t_{travel})} + 
\nn \\
&& \frac{2W_E}{W_E - W_T} \left( \rho_5(t_{travel}) - \frac{W_T}{W_E - W_T}\, \rho_1(t_{travel}) \right)\, e^{-(W_E+W_T)(t-t_{travel})} -
\nn \\
&& \frac{2W_E}{W_E - W_T}(\rho_5(t_{travel}) + \rho_1(t_{travel}))
\, e^{-W_T(t-t_{travel})} +
\nn \\
& & \frac{W^2_E}{(W_E - W_T)^2}\, \rho_1(t_{travel}) \, e^{-2W_T(t-t_{travel})} + 
\nn \\
&& \rho_{13}(t_{travel}) + 2\,\rho_5(t_{travel})\, + \rho_1(t_{travel}) 
\\
\rho_{34}(t) & = & 
\left( \rho_{14}(t_{travel}) - \frac{2W_T}{W_E - W_T}\, \rho_6(t_{travel})
+ \frac{W^2_T}{(W_E - W_T)^2}\, \rho_2(t_{travel}) \right)\, e^{-2W_E(t-t_{travel})} +  
\nn \\ 
& & \left( -2\, \rho_{14}(t_{travel})  - \frac{2(W_E - 2W_T)}{W_E - W_T}\, 
\rho_6(t_{travel}) + \frac{2 W_T}{W_E - W_T}\, \rho_2(t_{travel}) \right)\, 
e^{-W_E(t-t_{travel})} + 
\nn \\
&& \frac{2W_E}{W_E - W_T} \left( \rho_6(t_{travel}) - \frac{W_T}{W_E - W_T}\, \rho_2(t_{travel}) \right)\, e^{-(W_E+W_T)(t-t_{travel})} -
\nn \\ 
&& \frac{2W_E}{W_E - W_T}(\rho_6(t_{travel}) + \rho_2(t_{travel}))
\, e^{-W_T(t-t_{travel})} +
\nn \\
& & \frac{W^2_E}{(W_E - W_T)^2}\, \rho_2(t_{travel}) \, e^{-2W_T(t-t_{travel})} +
\nn \\
&& \rho_{14}(t_{travel}) +\, 2\,\rho_6(t_{travel}) +\, \rho_2(t_{travel}). 
\eea
\label{eq:rho3334}
\es

\begin{figure}[h]
\centerline{\epsfig{figure=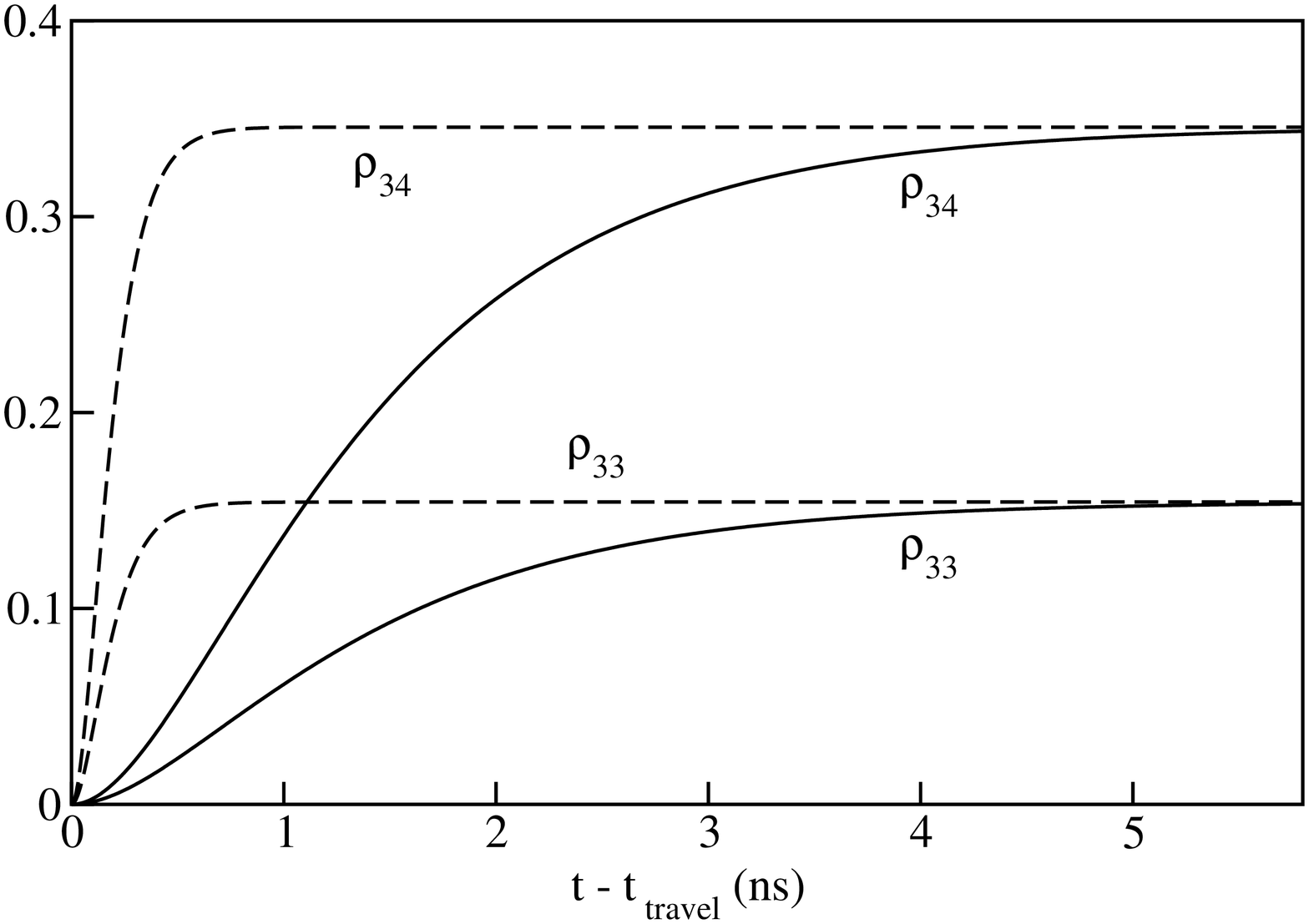,height=7.cm,width=8.cm}}
\caption[]{The probabilities $\rho_{33}$ to measure two spin-up electrons and $\rho_{34}$ to measure a spin-up and a spin-down electron in the left and right exits resp. for times $t\geq t_{travel}$. Parameters used are $\theta_1=\pi/2$, $\theta_2=\pi/8$ [so that $\rho_1(t_{ESR})$=0.154 and $\rho_2(t_{ESR})$=0.346], $t_{travel}-t_{ESR}=0.1$ ns, $W_E = 10^{10}$ s$^{-1}$ and
$W_T = 10^9 s^{-1} (9.9\, 10^9$ s$^{-1})$ for the solid (dashed) curves.
} 
\label{fig:rho3334}
\end{figure}

One can see directly from Eqs.~(\ref{eq:rho3334}) that the time dependence of $\rho_{33}$ and $\rho_{34}$ is determined by five exponential functions, whose relative magnitude depends on the ratio between $W_E$ and $W_T$. This is illustrated in 
Fig.~\ref{fig:rho3334}, which shows Eqs.~(\ref{eq:rho3334}) as a function of $t-t_{travel}$ for various rates $W_E$ and $W_T$. For $W_T \ll W_E$ the time needed to reach the stationary state (the average detection time) is dominated by the term $\sim e^{-W_T(t-t_{travel})}$, whereas for $W_T \approx W_E$ the terms
$\sim$ $e^{-2W_E(t-t_{travel})}$, $e^{-(W_E+W_T)(t-t_{travel})}$ and $e^{-2W_T(t-t_{travel})}$ dominate.

\section{Conclusion}
\label{sec-discussion}

In summary, we have presented an analytical solution of a set of coupled master equations that describe the time evolution of an entangled electron spin pair which can occupy 36 different quantum states in a double quantum dot nanostructure. Our method of solving these equations is based on separating the time evolution in three parts, namely two coherent rotations of the electron spins in the isolated quantum dots and the subsequent travel of the electrons through two quantum channels. As a result of this separation, the total number of master equations is split into various closed subsets of coupled equations. Our analytical solution is the first of its kind for a large set of coupled master equations and the same method can be used to study and predict the quantum evolution of other quantum systems which are described by a large set of quantum states. This type of analysis complements numerical approaches to study the dynamic evolution of complex quantum systems and allows to obtain qualitative insight in the competition between time scales in these systems.

\acknowledgments

Stimulating discussions with D.P. DiVincenzo are gratefully acknowledged. 
This work has been supported by the Stichting voor 
Fundamenteel Onderzoek der Materie (FOM), by the Netherlands Organisation for 
Scientific Research (NWO) and by the EU's Human Potential Research Network under 
contract No. HPRN-CT-2002-00309 (``QUACS''). 

\appendix
\section{Solution of Eqs.~(\ref{eq:masterequations1})-(\ref{eq:masterequations9})}
\label{app-A}

Using the substitution Eqs.~(\ref{eq:rhoESR}), Eqs.~(\ref{eq:masterequations1})-(\ref{eq:masterequations9}) transform into:
\begin{subequations}
\bea
\dot{x}_1 & = & - (W_{D \da \ua} + W_{D \ua \da} - \Gam) x_1 - \Del_L x_5 + W_{D \ua \da} e^{\Gam t} 
\label{eq:set2trans1}\\
\dot{x}_5 & = & \Del_L x_1 - \frac{\Del_L}{2} e^{\Gam t} 
\label{eq:set2trans2}\\
\dot{x}_2 & = & - (W_{D \da \ua} + W_{D \ua \da} - \Gam) x_2 - \del_L x_4 + W_{D \ua \da} e^{\Gam t} 
\label{eq:set2trans3}\\
\dot{x}_4 & = & \del_L x_2 - \frac{\del_L}{2} e^{\Gam t} 
\label{eq:set2trans4}\\
\dot{x}_3 & = & - (2 W_{D \da \ua} + 2 W_{D \ua \da} - \Gam) x_3 + \del_L x_6 + \Del_L x_7 + 2 W_{D \ua \da} e^{\Gam t}  + (W_{D \da \ua} - W_{D \ua \da}) (x_1 + x_2) 
\label{eq:set2trans5}\\
\dot{x}_6 & = & -  \del_L x_3 + 2 \Del_L x_8 + \frac{\del_L}{2} e^{\Gam t}  - 
\Del_L Z \\
\dot{x}_7 & = & -  \Del_L x_3 + 2 \del_L x_8 + \frac{\Del_L}{2} e^{\Gam t}  - 
\del_L Z \\
\dot{x}_8 & = & -  \frac{\Del_L}{2} x_6 - \frac{\del_L}{2} x_7,
\label{eq:set2trans8}
\eea
\label{eq:set2trans}
\end{subequations}
with $Z \equiv \mbox{\rm Re} [\til{\rho}_{1,4}(0) + \til{\rho}_{2,3}(0)]$.
In deriving Eqs.~(\ref{eq:set2trans}) we have used that
\bea 
\rho_4 & = & 1 - \rho_1 - \rho_2 - \rho_3 \nn \\
\mbox{\rm Re} \til{\rho}_{2,3} & = & -\mbox{\rm Re} \til{\rho}_{1,4} + Z\, e^{-\Gam t}. \nn
\label{eq:defC}
\eea
Equations.~(\ref{eq:set2trans}) consist of three sets of coupled equations, 
(\ref{eq:set2trans1})-(\ref{eq:set2trans2}), 
(\ref{eq:set2trans3})-(\ref{eq:set2trans4}) and 
(\ref{eq:set2trans5})-(\ref{eq:set2trans8}). 
The solution of the first two sets is given by:
\begin{subequations}
\bea
x_1 (t) & = & \left[ - \left( \frac{(W_{D\ua \da} + W_{D\da \ua} - \Gam) (x_1(0)-A_1) + 2 \Del_L (x_5(0)- A_2)}{2 \Omega_{\Del}} \right) 
\sin \Omega_{\Del}t\ + \right. \nn \\
&& \ \ \ \ \left. (x_1(0) - A_1) \cos \Omega_{\Del}t \right]\,    
e^{- \frac{1}{2} (W_{D\ua \da} + W_{D\da \ua} - \Gam)t} \ + \ 
A_1 e^{\Gam t} \\
x_5 (t) & = & \left[ \left( \frac{2 \Del_L (x_1(0)-A_1) + (W_{D\ua \da} + W_{D\da \ua} - \Gam) (x_5(0) - A_2)}{2 \Omega_{\Del}} \right) 
\sin \Omega_{\Del}t\ + \right. \nn \\
& & \ \ \ \ \left. (x_5(0) - A_2) \cos \Omega_{\Del}t \right] \,  
 e^{- \frac{1}{2} (W_{D\ua \da} + W_{D\da \ua} - \Gam)t} \ + \
A_2 e^{\Gam t} \\
x_2 (t) & = & \left[ - \left( \frac{(W_{D\ua \da} + W_{D\da \ua} - \Gam) (x_2(0)-A_3) + 2 \del_L (x_4(0) - A_4)}{2 \Omega_{\del}} \right) 
\sin \Omega_{\del}t\ + \right. \nn \\
&& \ \ \ \ \left. (x_2(0) - A_3) \cos \Omega_{\del}t \right]\,    
e^{- \frac{1}{2} (W_{D\ua \da} + W_{D\da \ua} - \Gam)t} \ + \ 
A_3 e^{\Gam t} \\
x_4 (t) & = & \left[ \left( \frac{2 \del_L (x_2(0)-A_3) + (W_{D\ua \da} + W_{D\da \ua} - \Gam) (x_4(0)-A_4)}{2 \Omega_{\del}} \right) 
\sin \Omega_{\del}t\ + \right. \nn \\
& & \ \ \ \ \left. (x_4(0) - A_4) \cos \Omega_{\del}t \right] \,  
 e^{- \frac{1}{2} (W_{D\ua \da} + W_{D\da \ua} - \Gam)t} \ + \
A_4 e^{\Gam t}
\eea
\label{eq:setx1}
\end{subequations}
with
\begin{subequations}
\bea
\Omega_{\Del} & = & \frac{1}{2} \sqrt{4 \Del_L^2 - (W_{D\ua \da} + W_{D\da \ua} - \Gam)^2} 
\label{eq:OmegaDel}\\
\Omega_{\del} & = & \frac{1}{2} \sqrt{4 \del_L^2 - (W_{D\ua \da} + W_{D\da \ua} - \Gam)^2} \\
A_1 & = & \frac{\Del_L^2 + 2 \Gam W_{D \ua \da}}{2 (\Del_L^2 + \Gam (W_{D\ua \da} + W_{D\da \ua}))} \\
A_2 & = & \frac{\Del_L (W_{D \ua \da} - W_{D \da \ua})}{2 (\Del_L^2 + \Gam (W_{D\ua \da} + W_{D\da \ua}))}
\\
A_3 & = & \frac{\del_L^2 + 2 \Gam W_{D\ua \da}}{2 (\del_L^2 + \Gam (W_{D\ua \da} + W_{D\da \ua}))} \\
A_4 & = & \frac{\del_L (W_{D\ua \da} - W_{D\da \ua})}{2 (\del_L^2 + \Gam (W_{D\ua \da} + W_{D\da \ua}))}. 
\eea
\end{subequations}
So far no approximations have been made, apart from assuming the decoherence rate $\Gam$ to be equal for all off-diagonal terms of the density matrix $\rho$ [Eqs.~(\ref{eq:masterequations})]. In order to obtain the solution of the remaining equations (\ref{eq:set2trans5})-(\ref{eq:set2trans8}) we assume $\del_L=0$ (no influence of the ESR field on the spin in the right dot) and $W_{D\ua \da} = W_{D\da \ua}=0$~\cite{general} and find~\cite{special}:
\begin{subequations}
\bea
x_3 (t) & = & \left[ \frac{ \Gam \left(x_3(0)- \frac{1}{2}\right) + 2 \Del_L x_7(0)}{2 \tilde{\Omega}_{\Del}}\, \sin \tilde{\Omega}_{\Del} t\ + (x_3(0) - \frac{1}{2}) \cos \tilde{\Omega}_{\Del} t \right]\,    
e^{\frac{\Gam}{2} t} \ + \ 
\frac{1}{2} e^{\Gam t} \\
x_6 (t) & = & x_6(0) \cos \Del_L t + (2 x_8(0) - Z) \sin \Del_L t \\
x_7 (t) & = & \left[ \frac{ - 2 \Del_L \left( x_3(0)-\frac{1}{2} \right) - \Gam x_7(0)}{2 \tilde{\Omega}_{\Del}}\, \sin \tilde{\Omega}_{\Del} t\ + 
x_7(0)  \cos \tilde{\Omega}_{\Del} t \right]\,    
e^{\frac{\Gam}{2} t} \\
x_8 (t) & = & \frac{1}{2}\left[ - x_6(0) \sin \Del_L t + (2x_8(0) - Z) \cos \Del_L t 
+ Z \right],
\eea
\label{eq:setx2}
\end{subequations}
with $\tilde{\Omega}_{\Del} = \frac{1}{2} \sqrt{4 \Del_L^2 - \Gam^2}$.

\section{Coefficients of Eqs.~(\ref{eq:sol13to16}) and (\ref{eq:sol25to32})}
\label{app-B}

The coefficients in Eqs.~(\ref{eq:sol13to16}) are given by:
\bea
E & = & \frac{-\rho_{13}(t_{ESR}) + \rho_{16}(t_{ESR}) + H_{13} - H_{16} + K_{13} - K_{16} + L_{13} - L_{16}}{2} \nn \\
\tilde{E} & = & \frac{\rho_{14}(t_{ESR}) - \rho_{15}(t_{ESR}) - H_{14} + H_{15} - K_{14} + K_{15} - L_{14} + L_{15}}{2} \nn \\
F & = & \frac{\rho_{13}(t_{ESR}) + \rho_{14}(t_{ESR}) + \rho_{15}(t_{ESR}) +  \rho_{16}(t_{ESR}) - 2 (K_{14} + K_{15}) - L_{13} - L_{14} - L_{15} - L_{16}}{2} \nn \\
\tilde{F} & = & \frac{-\rho_{13}(t_{ESR}) + \rho_{14}(t_{ESR}) + \rho_{15}(t_{ESR}) -  \rho_{16}(t_{ESR}) - 2 (H_{14} + H_{15}) + L_{13} - L_{14} - L_{15} + L_{16}}{2} \nn \\
H_{13} & = & \frac{W_T[(W_E-W_T+W_C)(A_{5,7,1,3}+A_{9,10,1,2}) + W_C(A_{6,8,2,4} + A_{11,12,3,4})]}{(W_E-W_T)(W_E-W_T+2W_C)} \nn \\
H_{14} & = & \frac{W_T[W_C (A_{5,7,1,3} - A_{11,12,3,4}) + (W_E-W_T+W_C)(A_{6,8,2,4} - A_{9,10,1,2})]}{(W_E-W_T)(W_E-W_T+2W_C)} \nn \\
H_{15} & = & - H_{14} [(5,7,1,3) \leftrightarrow (6,8,2,4), (9,10,1,2) \leftrightarrow (11,12,3,4)] \nn \\
H_{16} & = & - H_{13} [(5,7,1,3) \leftrightarrow (6,8,2,4), (9,10,1,2) \leftrightarrow (11,12,3,4)] \nn \\
K_{13} & = & H_{13} (A \rightarrow B) \\
K_{14} & = & \frac{W_T[W_C (B_{5,7,1,3} + B_{11,12,3,4}) + (W_E-W_T+W_C)(B_{6,8,2,4} + B_{9,10,1,2})]}{(W_E-W_T)(W_E-W_T+2W_C)} \nn \\ 
K_{15} & = & K_{14} [(5,7,1,3) \leftrightarrow (6,8,2,4), (9,10,1,2) \leftrightarrow (11,12,3,4)] \nn \\
K_{16} & = & H_{16} (A \rightarrow -B) \nn \\
L_{13} & = & W_T[(2(W_E-W_T+W_C)^2-W_C^2)(C_{1,2}+C_{1,3}) + W_C^2(D_{2,4}+D_{3,4}) + \nn \\ 
&& \hspace*{0.5cm} W_C(W_E-W_T+W_C)(C_{2,4}+C_{3,4}+D_{1,2}+D_{1,3})]/\nn \\
&& [4(W_E-W_T)(W_E-W_T+2W_C)(W_E-W_T+W_C)] \nn \\
L_{14} & = & W_T[W_C(W_E-W_T+W_C)(C_{1,2}+C_{1,3}+D_{2,4}+D_{3,4}) +
2(W_E-W_T+W_C)^2(C_{2,4}+D_{1,2})  \nn \\
&& + W_C^2(C_{3,4}-C_{2,4}-D_{1,2}+D_{1,3})]/[4(W_E-W_T)(W_E-W_T+2W_C)(W_E-W_T+W_C)] \nn \\
L_{15} & = & L_{14} (C_{2,4} \leftrightarrow C_{3,4}\, , D_{1,2} \leftrightarrow D_{1,3}) \nn \\
L_{16} & = & L_{13} (C_{1,2} \leftrightarrow D_{2,4}\, , C_{1,3} \leftrightarrow D_{3,4}). \nn
\eea

For (i,j)=(25,26) the coefficients in Eqs.~(\ref{eq:sol25to32}) are given by
\bea
P_{25,26} & = & P_{26,25} = - \frac{1}{2}\, \sum_{k=1}^{7} (M_{25,k} + M_{26,k}) \nn \\
Q_{25,26} & = & - Q_{26,25} = - \frac{1}{2}\, \sum_{k=1}^{7} (M_{25,k} - M_{26,k})\nn \\
M_{25,1} & = & \frac{(W_E + W_C)E + W_C \tilde{E}}{W_E + 2 W_C}\nn \\
M_{26,1} & = & M_{25,1} (E \leftrightarrow - \tilde{E})\nn \\
M_{25,2} & = & M_{26,2} = - \frac{1}{2} F\nn \\
M_{25,3} & = & - M_{26,3} = \frac{W_E}{2(W_E + 2W_C)} \tilde{F} \nn \\
M_{25,4} & = & \frac{W_E[-(W_T+W_C)((W_E+2W_C)H_{13} - W_T A_{5,7,1,3}) + W_C((W_E+2W_C)H_{14}-W_T A_{6,8,2,4})]}{W_T(W_E+2W_C)(W_T+2W_C)}\nn \\
M_{26,4} & = & M_{25,4}(H_{13} \leftrightarrow H_{14}, A_{5,7,1,3} \leftrightarrow 
A_{6,8,2,4}) \label{eq:coef25to32} \\
M_{25,5} & = & \frac{(W_C-W_T)(W_E K_{13}- W_T B_{5,7,1,3}) + W_C (W_E K_{14}- W_T B_{6,8,2,4})}{W_T(W_T-2W_C)} \nn \\
M_{26,5} & = & M_{25,5}(K_{13} \leftrightarrow K_{14}, B_{5,7,1,3} \leftrightarrow 
B_{6,8,2,4}) \nn \\
M_{25,6} & = & \frac{W_E [(W_E-2W_T+W_C)(L_{13}-C_{1,3}) + W_C (L_{14}-C_{2,4})]}{(W_E- 2 W_T+2W_C)(W_E-2W_T)}\nn \\
M_{26,6} & = & M_{25,6}(L_{13} \leftrightarrow L_{14}, C_{1,3} \leftrightarrow 
C_{2,4}) \nn \\
M_{25,7} & = & W_T \left[(W_E-W_T+W_C)[- A_{17} + B_{17} + C_{17}] + W_C [- A_{18} + B_{18} + C_{18}] \right]/ \nn \\
& & \left[ (W_E- W_T+2 W_C)(W_E-W_T) \right]\nn \\
M_{26,7} & = & M_{25,7}(17 \leftrightarrow 18). \nn
\eea

The coefficients in Eqs.~(\ref{eq:sol25to32}) for (i,j)=(27,28), (29,31) and (30,32) are obtained from Eqs.~(\ref{eq:coef25to32}) by replacing indices as given in Table~\ref{table:indices}.
\newpage
\begin{table}
\begin{tabular}{|c|c|c|}
\hline \hline
(27,28) & (29,31)& (30,32) \\ \hline \hline
25 $\rightarrow$ 28 & 25 $\rightarrow$ 29 & 25 $\rightarrow$ 32 \\ \hline
26 $\rightarrow$ 27 & 26 $\rightarrow$ 31 & 26 $\rightarrow$ 30 \\ \hline
13 $\rightarrow$ 16 & & 13 $\rightarrow$ 16 \\ \hline
14 $\rightarrow$ 15 & 14 $\rightarrow$ 15 & \\ \hline
17 $\rightarrow$ 20 & 17 $\rightarrow$ 21 & 17 $\rightarrow$ 24 \\ \hline
18 $\rightarrow$ 19 & 18 $\rightarrow$ 23 & 18 $\rightarrow$ 22 \\ \hline
$E \leftrightarrow -E$ & & $E \leftrightarrow - E$
\\ \hline
$\tilde{E} \leftrightarrow - \tilde{E}$ & $\tilde{E} \leftrightarrow - \tilde{E}$ & 
\\ \hline
$A_{5,7,1,3} \leftrightarrow - A_{6,8,2,4}$ & (5,7,1,3) $\rightarrow$ (9,10,1,2) & 
$A_{5,7,1,3} \rightarrow - A_{11,12,3,4}$ 
\\ \hline
$B_{5,7,1,3} \leftrightarrow B_{6,8,2,4}$ & & $B_{5,7,1,3} \rightarrow B_{11,12,3,4}$ 
\\ \hline 
& (6,8,2,4) $\rightarrow$ (11,12,3,4) & $A_{6,8,2,4} \rightarrow - A_{9,10,1,2}$ 
\\ \hline
& & $B_{6,8,2,4} \rightarrow B_{9,10,1,2}$ 
\\ \hline
$C_{1,3} \rightarrow D_{2,4}$ & $C_{1,3} \rightarrow C_{1,2}$ & $C_{1,3} \rightarrow D_{3,4}$ 
\\ \hline
$C_{2,4} \rightarrow D_{1,3}$ & $C_{2,4} \rightarrow C_{3,4}$ & $C_{2,4} \rightarrow D_{1,2}$.
\\ \hline \hline
\end{tabular}
\caption{Required substitution of indices and coefficients in Eqs.~(\ref{eq:sol25to32}) in order to obtain the corresponding coefficients for $\rho_i(t)$\&$\rho_j(t)$ with $(i,j)\in \{(27,28), (29,31), (30,32)\}$}.
\label{table:indices}
\end{table}


\begin{thebibliography}{99}
\bibitem{dec}
Assuming the same decoherence rate $\Gam$ for both dots is not a crucial assumption 
and Eqs.~(\ref{eq:masterequations}) can straightforwardly
be solved for different decoherence rates in the left and right dot. The resulting solution is qualitatively of the same form as Eqs.~(\ref{eq:set8}) and (\ref{eq:rhoESR}), with only more lengthy expressions.
\bibitem{RWA}
The RWA approximation~\cite{meys99} is not essential and one can solve Eqs.~(\ref{eq:masterequations}) without it, which results in a solution that is qualitatively similar. Since
$\omega t_1 \ll 1$ in our system~\cite{blaa05}, where $t_1$ is the time required to rotate one spin, 
the RWA is an excellent approximation in practice.
\bibitem{general}
For the general case the expressions are rather lengthy, but not qualitatively different.
\bibitem{special}
The solutions for $x_6(t)$ and $x_8(t)$ given here are valid for $\delta_L=0$,
i.e. assuming the W's to be zero is not necessary.
\end{thebibliography}
\end{document}